\documentclass[final]{svjour3}
\usepackage{graphicx}
\usepackage{rotating}
\usepackage{amssymb}
\usepackage{mathptmx}
\usepackage{gensymb}
\usepackage[square,sort,comma,numbers]{natbib}
\usepackage[dvipsnames]{xcolor}
\setcitestyle{square}
\makeatletter
\journalname{Journal of Low Temperature Physics}

\begin{document}

\newcommand{\hdblarrow}{H\makebox[0.9ex][l]{$\downdownarrows$}-}
\title{Developing AlMn films for Argonne TES fabrication}

\author{E.M. Vavagiakis$^1$, N.F. Cothard$^2$, J.R. Stevens$^1$, C.L. Chang$^3$, M.D. Niemack$^1$, G. Wang$^3$, V.G. Yefremenko$^3$, J. Zhang$^3$}

\institute{$^1$Department of Physics, Cornell University, Ithaca, NY 14853, USA\\ 
$^2$Department of Applied Physics, Cornell University, Ithaca, NY 14853, USA\\
$^3$Argonne National laboratory, 9700 S Cass Ave, Lemont, IL 60439, USA\\
Tel.: (607) 255-0474\\
\email{ev66@cornell.edu}}

\titlerunning{Developing AlMn films for Argonne TES fabrication}
\authorrunning{E.M. Vavagiakis et al.}

\maketitle

\begin{abstract}

The reference design for the next-generation cosmic microwave background (CMB) experiment, CMB-S4, relies on large arrays of transition edge sensor (TES) bolometers coupled to Superconducting Quantum Interference Device (SQUID)-based readout systems. Mapping the CMB to near cosmic variance limits will enable the search for signatures of inflation and constrain dark energy and neutrino physics. AlMn TESes provide simple film manufacturing and highly uniform arrays over large areas to meet the requirements of the CMB-S4 experiment. TES parameters such as critical temperature and normal resistance must be tuned to experiment specifications and can be varied based on geometry and steps in the fabrication process such as deposition layering, geometry, and baking time and temperature. Using four-terminal sensing, we measured $T_C$ and $R_N$ of AlMn 2000 ppm films and devices of varying thicknesses fabricated at Argonne National Laboratory to motivate device geometries and fabrication processes to tune $T_C$ to 150-200 mK and $R_N$ to $\sim$10 mOhms. Measurements of IV curves and time constants for the resulting devices of varying leg length were made using time-division SQUID multiplexing, and determined $T_C$, $G$, $k$, $f_{3db}$, and $R_N$. We present the results of these tests along with the geometries and fabrication steps used to tune the device parameters to the desired limits.


\end{abstract}

\section{Introduction}

The cosmic microwave background (CMB) contains a wealth of information about the origins, evolution, and fundamental physics of our universe. CMB-S4 is a proposed ``Stage-4" ground-based CMB experiment that will map the polarization of the CMB in multiple frequency bands to nearly the cosmic variance limit. CMB-S4 targets science goals that include characterizing dark energy and dark matter, searching for signatures of inflation in the early universe, measuring the sum of the neutrino masses, and mapping the universe's matter distribution \cite{CMBS4Science}.

CMB-S4 will rely on large arrays of Transition-Edge Sensors (TESes) coupled to Superconducting Quantum Interference Device (SQUID)-based readout systems to reach its sensitivity targets \cite{CMBS4Technology,CMBS4CD}. TESes are a mature technology and have been thoroughly demonstrated to achieve CMB science goals. TESes are also scalable, enabling the order of magnitude higher detector counts required by CMB-S4. The choice of AlMn TESes will provide CMB-S4 with simple film manufacturing and highly uniform distributions of device parameters over large arrays \cite{Li2016,Ho2017}.

Monolithic arrays of multichroic TESes have been previously fabricated at Argonne National Laboratory and deployed in the South Pole Telescope's SPT-3G camera \cite{Posada2015}. These Ti/Au bolometers have a critical temperature of $T_C$ = 420 mK and a normal resistance of $R_N$ = 2 $\ohm$. To meet the experiment design requirements for CMB-S4, AlMn$_{2000 ppm}$ TES fabrication will target a $T_C$ $\sim$ 150-200 mK and an $R_N$ $\sim$ 10-20 m$\ohm$. While SPT-3G used a $^4$He-$^3$He-$^3$He absorption refrigerator with a bath temperature of $\sim$250 mK, CMB-S4 will use dilution refrigerators with bath temperatures \textless100 mK \cite{Sobrin}. Reducing the bath temperature will reduce the thermal fluctuation noise. Reducing $R_N$ will optimize the detector arrays for TDM or microwave-SQUID multiplexing ($\mu$MUX) readout. The fabrication process at Argonne will therefore need to be tuned to achieve these device parameters. Variables in the fabrication process such as the Mn concentration, thickness and geometry of films, thermal annealing temperature, and presence of additional material layers all affect the AlMn device parameters and must be tuned to meet specifications through repeated rounds of fabrication and testing \cite{Schmidt,Li2016}. 

\section{Fabrication}

Film samples and devices were fabricated at Argonne National Laboratory and tested at Cornell University to develop the fabrication process. Bare film studies were performed to study the impact of substrate choice and Ti (or Mo) buffer layer between AlMn and Au depositions. A film stack of 2000ppm AlMn on SiO2 with a Ti15nm/Au15nm top layer was chosen based on good critical temperature repeatability and was derived from the Ti/Au SPT-3G array fabrication (Fig. 1). The top layer of Au was chosen to prevent oxidation and protect the underlying films, while remaining conductive to connect leads to the device. Patterned films were then tested to predict device $T_C$ and $R_N$.

\begin{figure}[htbp]
\begin{centering}
\includegraphics[width=0.4\linewidth]{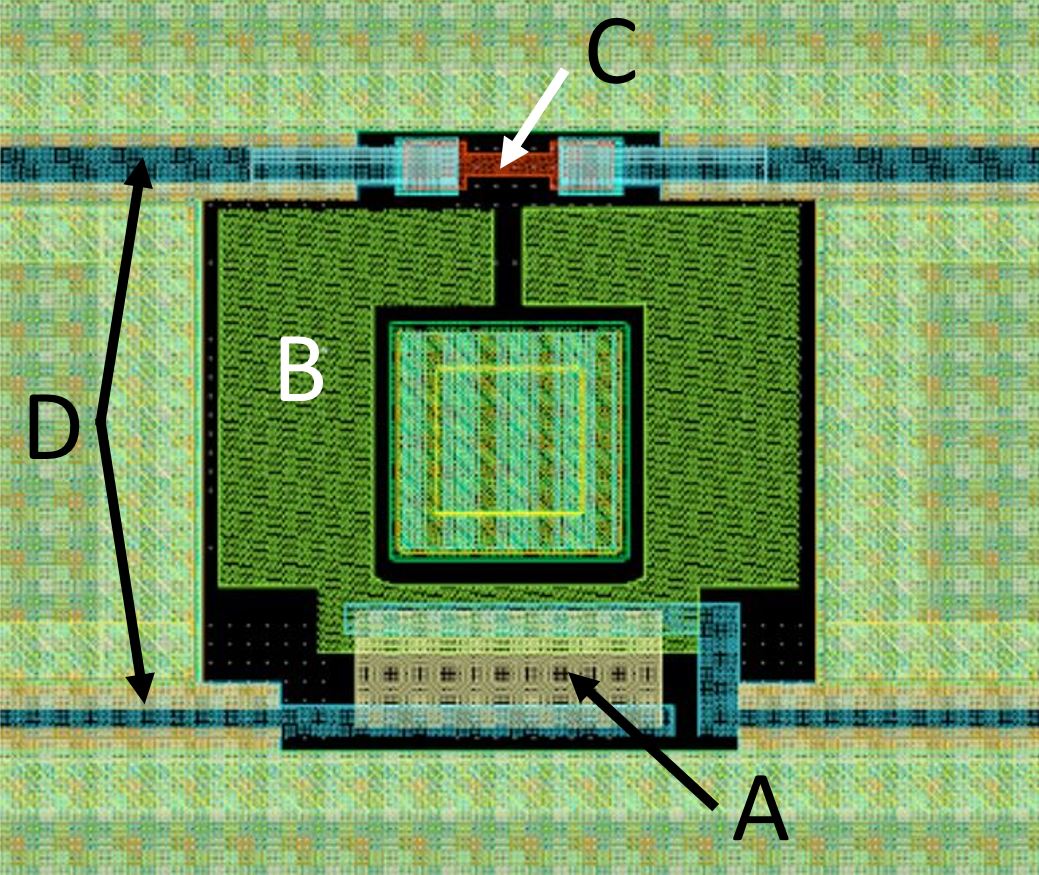}
\caption{A modified SPT-3G TES island, where (A) is the AlMn TES film, (B) is Pd for heat capacity stabilization, (C) is a load resistor, and (D) are the Nb leads. The AlMn200nm/Ti15nm/Au15nm TES film is 15 $\mu$m long by 80 $\mu$m wide.}
\end{centering}
\end{figure}

\noindent The thickness and aspect ratio of the film stacks was then iterated upon to target the desired $R_N$. Baking time and temperature were selected to target $T_C$. A lift-off process is used for TES patterns, which can potentially lead to tapered edges and interaction between the Al and Au layers, and thus to poor transitions for thicker films. Furthermore, direct contact of Al and Au leads to the formation of intermetallic compounds and the variation of superconducting properties. Such effects as well as proximity effects were considered while iterating on patterned films and TESes. 

\section{Testing}

Film samples, patterned films, and TESes from Argonne were tested at Argonne and Cornell University. 

\subsection{Four-lead measurements}

Samples were wire bonded and affixed with rubber cement to a printed circuit board (PCB) and mounted to the coldest (100 mK) stage of a dilution refrigerator with an internal 300 K magnetic shield. Four-lead measurements were taken of the samples, which precisely measure low resistance values by eliminating the lead and contact resistances from the measurements. Temperature was varied as resistances were logged via a Lakeshore AC resistance bridge, reading out the superconducting transition and measuring $T_C$ and $R_N$. For each transition, $T_C$ was taken to be the temperature value at 50$\%$ $R_N$, where $R_N$ is the resistance value measured at 2 mK above the last superconducting data point in the resistance versus temperature curve (Fig. 2). Excitation currents were varied to ensure their choice didn't significantly affect the measured $T_C$. Based on the noise in the obtained curves due to temperature fluctuations and Lakeshore measurements, the total uncertainty on $T_C$ is taken to be 2 mK, and the uncertainty on $R_N$ to be 0.2 $m\ohm$. 

\begin{figure}[htbp]
\begin{centering}
\includegraphics[width=0.8\linewidth]{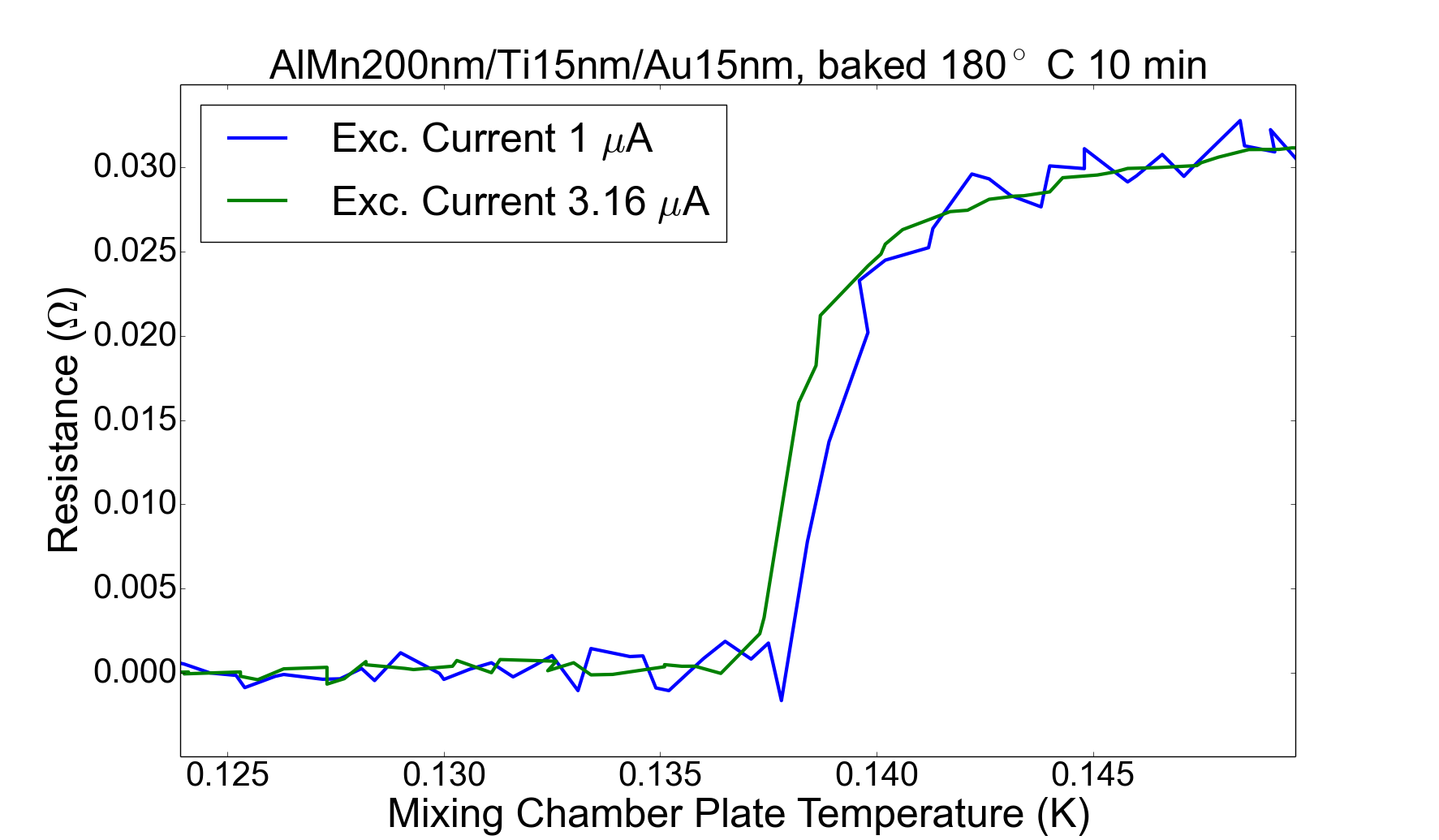}
\caption{Example of the four-lead R vs. T measurement of a film superconducting transition. Choice of excitation current influences results, and $T_C$ is recorded for an excitation current which has been reduced enough to keep $T_C$ consistent within 2 mK.}
\end{centering}
\end{figure}

\subsection{TDM SQUID readout measurements}

Fabricated TESes of two different leg lengths (865 $\mu$m and 446 $\mu$m) of equal cross-sectional area (21 $\mu$m wide, 1$\mu$m SiN + 0.3 $\mu$m Nb + 0.5 $\mu$m SIO2 thick),  were mounted to a PCB on the cold stage of the dilution refrigerator, and read out using the same time division multiplexing (TDM) system used in AdvACT with NIST SQUIDs \cite{Henderson2016}. TDM measurements of current-voltage (IV) curves at various bath temperatures were aquired for a ``long" and ``short" leg length AlMn$_{2000ppm}$200nm/Ti15nm/Au15nm TES. We define $P_{sat}$ to be the bias power which drives the TES to 90$\%$ of its normal resistance, $R_N$. The measured $P_{sat}$ values are fit to the model

\begin{equation}
    P_{sat} = K(T_C^n - T_{bath}^n),
\end{equation}

\noindent where $P_{sat}$ is related to the bath temperature $T_{bath}$ and the critical temperature ($T_C$) of the device. Because $K$ is degenerate with $n$ in our fits, we hold $n$ fixed at a range of values while fitting the other parameters, and finally determine the best fit $n$ by plotting $n$ vs $\chi^2$ and obtaining the minimum $\chi^2$. The thermal conductivity, $G$, of the device is then given by

\begin{equation}
    G=\frac{dP_{sat}}{dT_C}=nKT_C^{n-1}.
\end{equation}

\noindent The temporal response of a given TES can be defined in terms of $f_{3dB} = 1/2\pi\tau_{eff}$, where $\tau_{eff}$ is the effective time thermal constant of the TES operated under negative electrothermal feedback. To measure $f_{3dB}$, a small amplitude square wave is added to the DC detector voltage bias. The detector response is sampled quickly at $\sim$1600 Hz. The time constant, $\tau_{eff}$, is extracted by fitting the response to a single pole exponential. These measurements are performed at multiple bath temperatures and at several different points on the superconducting transition \cite{Koopman2017}. As a next step, we are working on noise data acquisitions for Argonne TESes \cite{Stevens2019}.

\section{Results}

\subsection{Four-lead measurement results}

The critical temperature and normal resistance measurements for TESes and films varying in material, geometry, and bake time and temperature are given in Table 1. Baking was found to shift $T_C$, and the films showed evidence of annealing, where $T_C$ doesn't change again after the first bake. Thicker and wider geometries lowered $R_N$ to approach the target value, and aspect ratios were also found to influence $T_C$ (Fig. 3).

\begin{table}
\begin{center}
\begin{tabular}{ |c|c|c|c| } 
 \hline
 \textbf{Film} & \textbf{Heating $\degree$C} & \textbf{$T_C$ [mK]} & \textbf{$R_N$ [$\ohm$]}\\ 
 \hline
 Ti5nm/AlMn$_{1200ppm}$100nm/Ti5nm/Au20nm & 0 & 196 & 0.86\\
 Ti5nm/AlMn$_{1200ppm}$100nm/Ti5nm/Au20nm & 180/10 min & 153 & 0.80\\
 Ti5nm/AlMn$_{1200ppm}$100nm/Ti5nm/Au20nm & 180/10 min,90/2 min,110/2 min &141 & 0.77\\
 \hline
 AlMn$_{1200ppm}$100nm/Mo10nm/Au20nm & 0 & 371 & 0.27\\
 AlMn$_{1200ppm}$100nm/Mo10nm/Au20nm & 180/10 min & 399 & 0.37\\
 \hline
 AlMn$_{2000ppm}$100nm/Ti15nm/Au15nm & 0 & 223 & 0.43\\
 AlMn$_{2000ppm}$100nm/Ti15nm/Au15nm & 0 & 159 & 1.65\\
 AlMn$_{2000ppm}$100nm/Ti15nm/Au15nm & 180/10 min & 181 & 1.45\\
 AlMn$_{2000ppm}$100nm/Ti15nm/Au15nm & 180/10 min,90/2 min, 110/1.5 min & 180 & 1.46\\
 AlMn$_{2000ppm}$100nm/Ti15nm/Au15nm & 0 & 179 & 0.09\\
 AlMn$_{2000ppm}$100nm/Ti15nm/Au15nm & 180/10 min & 193 & 0.05\\
 AlMn$_{2000ppm}$100nm/Ti15nm/Au15nm & 180/10 min,90/2 min, 110/1.5 min & 172 & 0.05\\
 AlMn$_{2000ppm}$100nm/Ti15nm/Au15nm & 0 & 167 & 3.31\\
 AlMn$_{2000ppm}$100nm/Ti15nm/Au15nm & 180/10 min & 160 & 3.06\\
 \hline
 AlMn$_{2000ppm}$80nm & 0 & 335 & 2.10\\
 \hline
 AlMn$_{2000ppm}$550nm/Ti15nm/Au15nm & 0 & 151 & 0.22\\
 AlMn$_{2000ppm}$550nm/Ti15nm/Au15nm & 0 & 146 & 0.11\\
 AlMn$_{2000ppm}$550nm/Ti15nm/Au15nm & 180/10 min & 66 & 0.11\\
 AlMn$_{2000ppm}$550nm/Ti15nm/Au15nm & 180/20 min & 67 & 0.11\\
 \hline
 AlMn$_{2000ppm}$200nm/Ti15nm/Au15nm* & 0 & 127 & 0.03\\
 AlMn$_{2000ppm}$200nm/Ti15nm/Au15nm* & 180/10 min & 129 & 0.04\\
 AlMn$_{2000ppm}$200nm/Ti15nm/Au15nm* & 180/10 min & 140 & 0.04\\
 \hline
 AlMn$_{2000ppm}$380nm/Ti15nm/Au15nm* & 180/10 min & 179 & 0.02\\
 AlMn$_{2000ppm}$550nm/Ti15nm/Au15nm* & 180/10 min & 167 & 0.02\\
 AlMn$_{2000ppm}$550nm/Ti15nm/Au15nm* & 0 & 219 & 0.02\\
 AlMn$_{2000ppm}$550nm/Ti15nm/Au15nm* & 180/20 sec + 10 min & 88 & 0.01\\
 AlMn$_{2000ppm}$550nm/Ti15nm/Au15nm* & 180/60 sec + 10 min & 86 & 0.01\\
 AlMn$_{2000ppm}$200nm/Ti15nm/Au15nm & 180 & 89 & 0.03\\
 AlMn$_{2000ppm}$200nm/Ti15nm/Au15nm & 210 & 125 & 0.03\\
 AlMn$_{2000ppm}$200nm/Ti15nm/Au15nm & 240 & 174 & 0.03\\
 \hline \hline
 \textbf{Film} & \textbf{Length [$\mu$m] x Width [$\mu$m]} & \textbf{$T_C$} & \textbf{$R_N$}\\
 \hline
 AlMn$_{2000ppm}$200nm/Ti15nm/Au15nm & 5x80 & 201 & 0.01\\
 AlMn$_{2000ppm}$200nm/Ti15nm/Au15nm & 20x200 & 126 & 0.01\\
 AlMn$_{2000ppm}$200nm/Ti15nm/Au15nm & 5x140 & 172 & 0.01\\
 AlMn$_{2000ppm}$200nm/Ti15nm/Au15nm & 10x200 & 146 & 0.01\\
\hline 
\end{tabular}
\caption{Samples tested using four-lead measurements, with various materials, geometries, bake times and temperatures, and resulting critical temperatures and normal resistances. *=TES devices.}
\vspace{-1.5em}
\end{center}
\label{table:table1}
\end{table}

\subsection{TDM measurement results}

The results of $P_{sat}$ fits (Fig. 4) for two Argonne devices are shown in Table 2. The results of bias step measurements are shown in Fig. 5. We observe a large decrease in time constant as the detectors are biased lower on the superconducting transition, which is consistent with previous AlMn TES measurements \cite{Koopman2017}.

\begin{table}
\begin{center}
\begin{tabular}{ |c|c|c| } 
 \hline
 \textbf{Parameter} & \textbf{Leg Length 865 $\mu$m Fit} & \textbf{Leg Length 446 $\mu$m Fit} \\ 
 \hline
 $T_C$ & 122 mK & 136 mK\\
 $G$ & 0.013 pW/mK & 0.024 pW/mK\\
 $k$ & 4.6e-7 pW/mK$^n$ & 1.5e-6 pW/mK$^n$\\
 $R_N$ & 21.9 m$\ohm$ & 21.7 m$\ohm$ \\
\hline 
\end{tabular}
\caption{Parameter fits to $P_{sat}$ measurements from Fig. 4, Eqs. 1, 2, for two Argonne devices of different leg length and identical leg cross-sectional area. The best fit values for these data were $n$ = 2.76 and $n$ = 2.91 for the short and long leg lengths respectively.}
\vspace{-1.5em}
\end{center}
\label{table:table2}
\end{table}

\begin{figure}[htbp]
\begin{centering}
\includegraphics[width=1.0\linewidth]{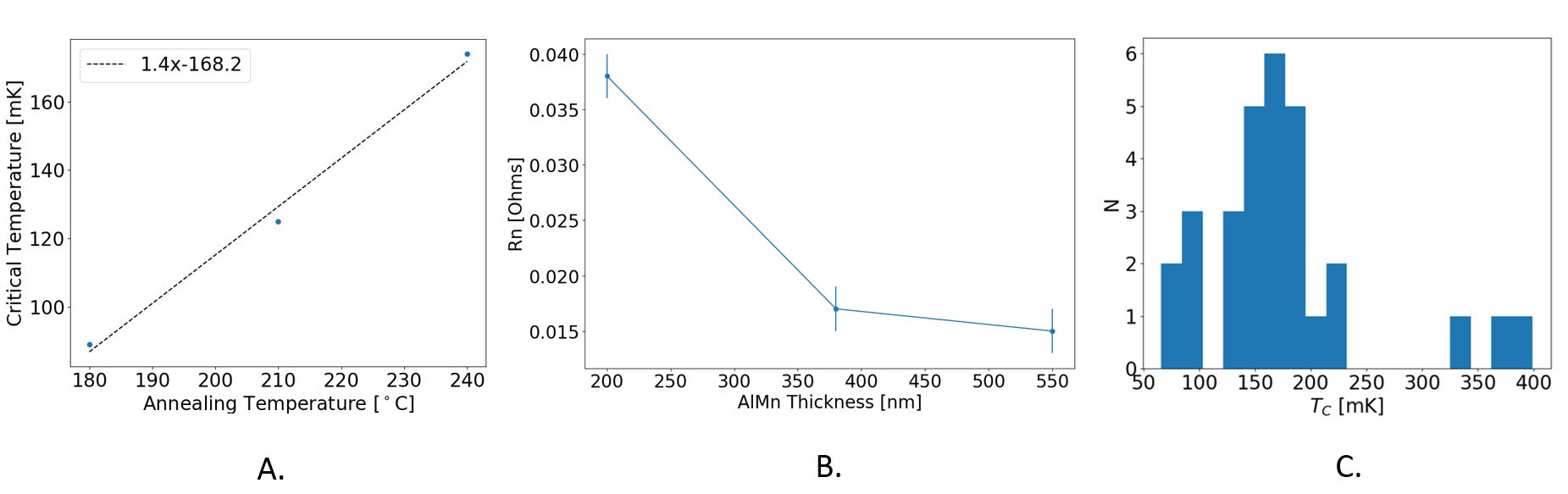}
\caption{\textit{A:} Critical temperature vs. annealing temperature for three AlMn$_{2000ppm}$200nm/Ti15nm/Au15nm films along with linear fit showing the effect of baking on $T_C$. \textit{B:} Normal resistance vs. AlMn thickness in nm for three film samples, showing $R_N$ decreasing with increasing material thickness. \textit{C:} Histogram of all Argonne sample critical temperatures measured at Cornell, highlighting a trend of narrowing in on the desired critical temperature of 150-200 mK for CMB-S4.}
\end{centering}
\end{figure}

\begin{figure}
    \centering
    \includegraphics[width=0.90\linewidth]{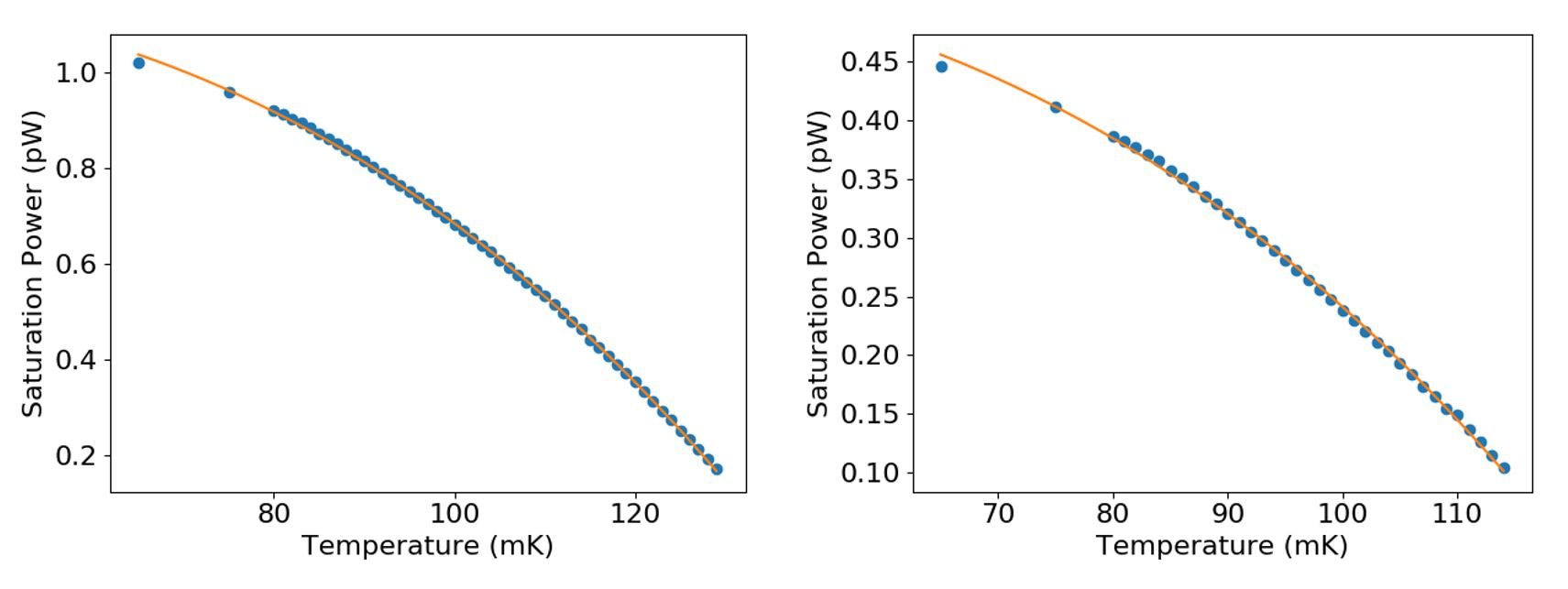}
    \caption{Saturation power versus temperature along with fits to $P_{sat}$ measurements for two devices. \textit{Left:} Argonne short leg length (446 $\mu$m) TES. \textit{Right:} Argonne long leg length (865 $\mu$m) TES. The best fit $n$ values for these data were 2.76 and 2.91 for the short and long leg lengths respectively.}
\end{figure}

\begin{figure}[htbp]
\begin{centering}
\includegraphics[width=1.0\linewidth]{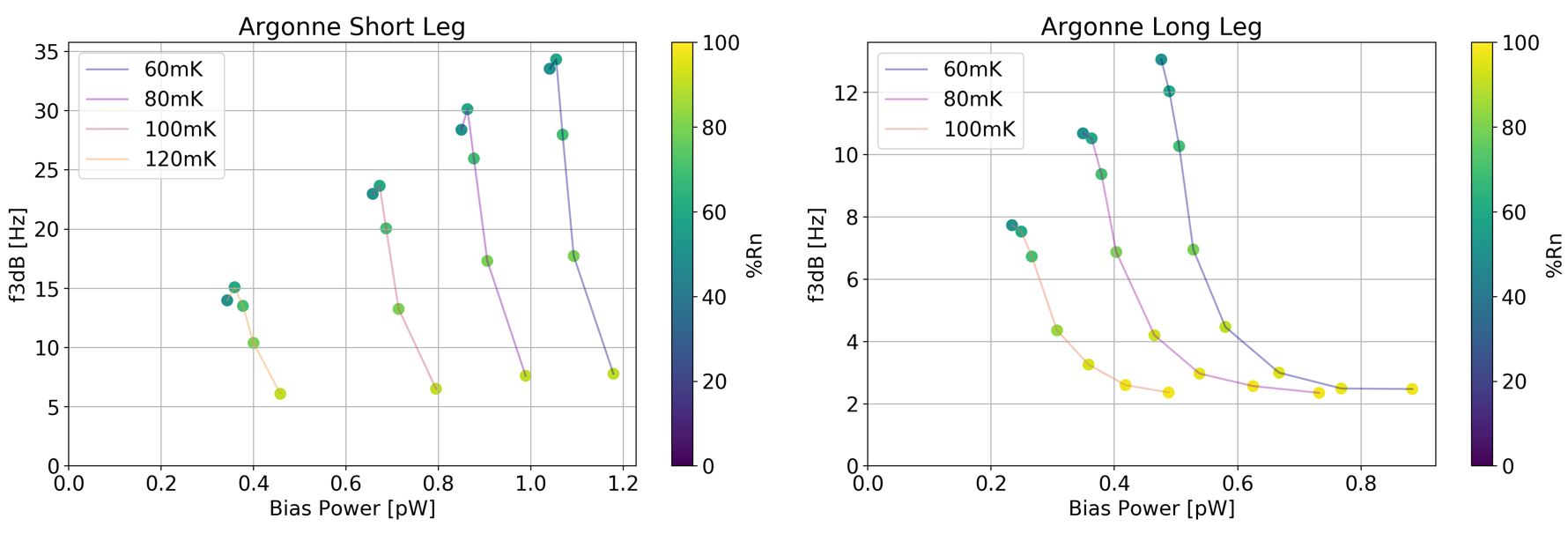}
\caption{Results of bias step measurements. $f_{3dB}$ versus bias power in $pW$ for four bath temperatures. \textit{Left:} Argonne short leg length (446 $\mu$m) TES. \textit{Right:} Argonne long leg length (865 $\mu$m) TES.}
\end{centering}
\end{figure}

\section{Summary}

We have measured $T_C$, $G$, $k$, f$_{3db}$, and $R_N$ for two Argonne AlMn TESes, and $T_C$ and $R_N$ for various film stacks to motivate the selection of materials, thickness, geometry, and bake time and temperature. Through iterative testing, we inform Argonne's fabrication process to produce TESes which have the desired critical temperature (150-200 mK) and normal resistance (10-20 m$\ohm$) to be used for precision measurements of the CMB in the next-generation CMB experiment, CMB-S4. Next steps will include fabrication of devices with higher saturation powers (2 pW-20 pW) and lower time constants (0.3-3 ms) to optimize the TES performance for all the CMB-S4 frequency bands.

\begin{acknowledgements}
Work at Argonne, including use of the Center for Nanoscale Materials, an Office of Science user facility, was supported by the U.S. Department of Energy, Office of Science, Office of Basic Energy Sciences and Office of High Energy Physics, under Contract No. DE-AC02-06CH11357. This material is based upon work supported by the National Science Foundation Graduate Research Fellowship Program under Grant No. DGE-1650441 (EMV) and NSF CAREER award No. 1454881 (MDN). Work by NFC was supported by a NASA Space Technology Research Fellowship.
\end{acknowledgements}

\end{document}